\documentclass[english,romanappendices]{IEEEtran}
\usepackage[T1]{fontenc}
\usepackage[latin9]{inputenc}
\usepackage{xcolor}
\usepackage{bm}
\usepackage{amsmath}
\usepackage{amsthm}
\usepackage{amssymb}
\usepackage{graphicx}

\makeatletter
\theoremstyle{plain}
\newtheorem{thm}{\protect\theoremname}
\theoremstyle{remark}
\newtheorem{rem}[thm]{\protect\remarkname}
\theoremstyle{plain}
\newtheorem{prop}[thm]{\protect\propositionname}
\theoremstyle{plain}
\newtheorem{lem}[thm]{\protect\lemmaname}

\PassOptionsToPackage{hyphens,spaces,obeyspaces}{url}
\usepackage{cite}
\usepackage{psfrag}
\usepackage{enumitem}
\usepackage[hyphens,spaces,obeyspaces]{url}
\setenumerate{label=(\roman*)}

\makeatother

\usepackage{babel}
\providecommand{\lemmaname}{Lemma}
\providecommand{\propositionname}{Proposition}
\providecommand{\remarkname}{Remark}
\providecommand{\theoremname}{Theorem}

\begin{document}

\title{Optimizing Reweighted Belief Propagation for Distributed Likelihood
Fusion Problems}

\author{Christopher Lindberg, Julien M. Hendrickx, and Henk Wymeersch\thanks{Christopher Lindberg is with Zenuity AB, Gothenburg, Sweden. email: christopher.lindberg@zenuity.com. Julien M. Hendrickx is with ICTEAM Institute/CORE, Université Catholique de Louvain, Louvain-la-Neuve, Belgium. email: julien.hendrickx@uclouvain.be. Henk Wymeersch is with the Electrical Engineering Department, Chalmers University of Technology, Gothenburg, Sweden. email: henkw@chalmers.se.
This work was supported, in part, by the European Research Council under Grant No. 258418 (COOPNET), and by the DYSCO (dynamical systems, control, and optimization) network, funded by the Interuniversity Attraction Poles Programme, initiated by the Belgian Federal Science Policy Office, and by the Concerted Research Action (ARC) of the French Community of Belgium.}}
\maketitle
\begin{abstract}
Belief propagation (BP) is a powerful tool to solve distributed inference
problems, though it is limited by short cycles in the corresponding
factor graph. Such cycles may lead to incorrect solutions or oscillatory
behavior. Only for certain types of problems are convergence properties
understood. We extend this knowledge by investigating the use of reweighted
BP for distributed likelihood fusion problems, which are characterized
by equality constraints along possibly short cycles. Through a linear
formulation of BP, we are able to analytically derive convergence
conditions for certain types of graphs and optimize the convergence
speed. We compare with standard belief consensus and observe significantly
faster convergence. 
\end{abstract}

\section{Introduction}

\IEEEPARstart{B}{elief} propagation (BP) \cite{pearl1986fusion,kschischang2001factor}
is a message-passing algorithm for approximate inference on graphs
of problems that arise in many different fields such as statistical
physics, computer vision, artificial intelligence, optimization, behavioral
modeling in social networks, and wireless communications \cite{yedidia2005constructing,sun2003stereo,moallemi2009convergence,chamley2013models}.
Examples of applications to wireless communications include detection
problems, localization and tracking, and decoding \cite{kabashima2003cdma,wymeersch2009cooperative,fossorier1999reduced,mceliece1998turbo}.
One of the more notable applications is iterative decoding algorithms
for capacity-approaching error-correcting codes, including LDPC and
turbo codes. Since BP is a message-passing algorithm, it is also suitable
for solving distributed problems in networks of cooperating nodes.
Examples include distributed cooperative decision making in cognitive
radio \cite{zarrin2008belief}, distributed cooperative localization
and/or tracking \cite{wymeersch2009cooperative,meyer2012simultaneous},
network synchronization \cite{etzlinger2014cooperative}, distributed
joint source channel decoding \cite{zhong2005ldgm}, and distributed
compressed sensing \cite{zhang2011belief}.

While BP generally works well in practice, convergence can in general
not be guaranteed. This phenomenon is especially apparent on graphs
that have cycles with strong interactions, with extreme case equality
constraints, which force variables to maintain the same value along
a cycle in the graph. An example of such a setting is the \emph{distributed
likelihood fusion problem}, where nodes in a network must agree on
a global likelihood function, based on locally available, mutually
independent observations. To mitigate the convergence issues for such
problems, one can apply a variation of BP \cite{wainwright2003tree,wainwright2005new,wymeersch2011uniformly,wymeersch2012uniformly,liu2012knowledge,liu2012low}
or apply methods from the field of distributed consensus \cite{olfati2007con,xiao2006distributed}.
In the first class, \cite{wainwright2003tree,wainwright2005new} introduced
the tree-reweighted BP (TRW-BP), which optimizes convex combinations
of cycle-free graphs (tree-graphs) to represent the original graph
problem, leading to promising performance at a cost of solving of
a high-dimensional optimization problem over spanning trees.  The
uniformly reweighted BP (URW-BP) algorithm \cite{wymeersch2011uniformly,wymeersch2012uniformly}
is a special case of TRW-BP that involves optimization over one parameter,
lending itself well for implementation in network settings, or where
computational efficiency is prioritized. URW-BP variations were applied
to improve decoding performance of LDPC codes in \cite{liu2012knowledge,liu2012low}.
In the second class, distributed likelihood fusion is solved using
distributed consensus methods, leading to approaches commonly termed
\emph{belief consensus}: \cite{olfati2007con} proposes a distributed
consensus method, whereby the convergence speed depends on a single
scalar parameter, which depends on the maximum node degree. A fast
version of such belief consensus was proposed in \cite{xiao2006distributed}
using Metropolis-type weights, which can be locally computed. However,
such consensus methods are generally slow on tree graphs for which
BP works well. 

In this paper, we cast URW-BP as a linear system (similar to the linear
BP expressions in \cite{dai2007consensus}), allowing eigen-analysis.
\textcolor{black}{Our contributions are summarized in three parts
as follows: (i) We show that for a certain class of network inference
problems (i.e., likelihood fusion problems) and certain network topologies
(i.e., trees, $k$-regular graphs, and variations of the latter),
both belief consensus and URW-BP can achieve convergence to the correct
beliefs; (ii) In such cases, we can analytically optimize the URW-BP
parameter to maximize the convergence rate, outperforming belief consensus;
(iii) As a side-result, we recover a new way to prove the finite-time
convergence of BP on trees. }

The remainder of the paper is organized as follows: In Section \ref{sec:sysmod}
we formalize the distributed likelihood fusion problem. Section \ref{sec:Algorithms}
introduces the algorithms which are used to solve the problem. Section
\ref{sec:Eigenvalues-and-Jordan} deals with the tools we use to analyze
the convergence behavior of these algorithms. In Section \ref{sec:Convergence-on-Tree-graphs},
we present the convergence analysis of the algorithms on tree graphs
and $k$-regular graphs respectively. In Section \ref{sec:Results-and-Discussion}
we present results from numerical simulations, and some discussion
of those. We conclude the paper in Section \ref{sec:Conclusion}. 

\subsection*{Notation}

We use boldface lowercase letters $\boldsymbol{x}$ for column vectors,
and boldface uppercase letters $\boldsymbol{X}$ for matrices. In
particular, $\boldsymbol{I}_{M}$ denotes an $M\times M$ identity
matrix, $\bm{O}_{M}$ denotes an $M\times M$ all zero matrix, $\boldsymbol{1}$
is the all one vector of appropriate size, and $\boldsymbol{0}$ is
the all zero vector of appropriate size. Sets are described by calligraphic
letters $\mathcal{X}$ and the cardinality of a set is denoted by
$|\mathcal{X}|$. The transpose of a vector is denoted by $[\cdot]^{\mathsf{T}}$.
The indicator function of a statement $\mathsf{P}$ is written as
$\mathbb{I}_{\left\{ \mathsf{P}\right\} }\in\{0,1\}$. We denote by
$\sum_{\sim x_{i}}f(\mathbf{x})$ the summation over all elements
in $\mathbf{x}$, except $x_{i}$. 

\section{Problem Formulation\label{sec:sysmod}}

We consider a network consisting of $N$ connected nodes which we
model by an undirected graph $\mathcal{G}=(\mathcal{V},\mathcal{E})$,
where $\mathcal{V}$ is the set of nodes and $\mathcal{E}$ is the
set of edges connecting the nodes. Associated with the graph $\mathcal{G}$
is the \emph{adjacency matrix }$\bm{A}$ with entries $A_{ij}=\mathbb{I}_{\left\{ (i,j)\in\mathcal{E}\right\} }$,
the \emph{degree matrix }$\bm{D}=\mathrm{diag}(\bm{A1})$, and the
\emph{Laplacian matrix }$\bm{L}=\bm{D}-\bm{A}$. For later use, let
$\mu_{1},\dots,\mu_{N}$ be the eigenvalues of $\bm{A}$ sorted such
that $\left|\mu_{1}\right|\ge\left|\mu_{2}\right|\ge\dots\ge\left|\mu_{N}\right|$.
We consider three types of graphs:
\begin{enumerate}
\item \emph{Tree-graphs: }The set of edges $\mathcal{E}$ connects all the
vertices (nodes) in $\mathcal{V}$ such that there are no cycles.
Nodes connected to exactly one node are called \emph{leaves}.
\item \emph{$k$-regular graphs: }All nodes are connected to exactly $k$
other nodes. 
\item \emph{General connected graphs:} There is no constraint on the edge
set, provided the graph is connected. 
\end{enumerate}
The aim of the network is to determine the posterior distribution
over a variable $\theta$ given independent local observations $y_{n}$,
at each node $n$. Hence, each node has access to a local likelihood
function $p(y_{n}|\theta)$ where the likelihood functions are conditionally
independent given $\theta$, and it is also assumed that each node
knows the prior distribution $p(\theta)$. The posterior distribution
can be factorized as
\begin{equation}
p(\theta|y_{1},\dots,y_{N})\propto p(\theta)\prod_{m=1}^{N}p(y_{m}|\theta),\label{eq:postfact}
\end{equation}
or equivalently in the log-domain as
\begin{equation}
\log p(\theta|y_{1},\dots,y_{N})\propto\log p(\theta)+\sum_{m=1}^{N}\log p(y_{m}|\theta).\label{eq:logfactor}
\end{equation}
We assume that $\theta$ is a discrete random variable that can only
take on $K$ distinct values.

\section{Two Solution Approaches\label{sec:Algorithms}}

In this section, we describe techniques that can be used to compute
the posterior distribution from the local likelihood functions at
each node in a distributed manner: belief consensus and belief propagation. 

\subsection{Belief Consensus}

The problem in (\ref{eq:logfactor}) can be solved by reaching \emph{consensus}
on the average of the log-likelihood functions, and multiplying the
consensus value by the number of nodes. The belief consensus algorithm
aims to compute the consensus value by letting the nodes iteratively
exchange information with their neighbors and updating their state
according to an update rule specified by the algorithm. Let the initial
state of the consensus algorithm\emph{ }of node $n$ be its local
likelihood function, i.e., $x_{n}^{(0)}(\theta)=\log p(y_{n}|\theta)$.
The network updating dynamics are described by
\begin{equation}
\bm{x}^{(\ell)}(\theta)=\bm{W}\bm{x}^{(\ell-1)}(\theta),
\end{equation}
where $\bm{W}$ is an appropriately chosen matrix, with $W_{nm}=0$
when $(m,n)\notin\mathcal{E}$. Examples include \emph{Metropolis
weighting}, where weighting is decided by all nodes determining its
outgoing weights and self-weight by
\begin{equation}
W_{nm}=\begin{cases}
1/\left(\max\left\{ \left|\mathcal{N}_{n}\right|,\left|\mathcal{N}_{m}\right|\right\} +1\right) & (n,m)\in\mathcal{E}\\
1-\sum_{u\in\mathcal{N}_{n}}W_{nu} & m=n\\
0 & \mathrm{otherwise}.
\end{cases}\label{eq:metropolis}
\end{equation}
or \emph{uniform-weight consensus}, where 

\begin{equation}
W_{nm}=\begin{cases}
\xi & (n,m)\in\mathcal{E}\\
1-\xi\left|\mathcal{N}_{n}\right| & m=n\\
0 & \mathrm{otherwise}.
\end{cases}\label{eq:maxdegree}
\end{equation}
If $\xi$ is chosen as $0<\xi<1/\max_{m}|\mathcal{N}_{m}|$ , then
$\bm{W}$ in either (\ref{eq:metropolis}) or (\ref{eq:maxdegree})
is a doubly stochastic matrix with one eigenvalue $1$ (with corresponding
normalized eigenvector $\bm{1}/\sqrt{N}$), while all other eigenvalues
are strictly smaller than 1 in absolute value. Hence, the convergence
rate of belief consensus is determined by the second largest eigenvalue
of $\bm{W}$. Moreover, it can be shown that for any node $n$
\begin{equation}
\lim_{\ell\rightarrow\infty}x_{n}^{(\ell)}(\theta)=\frac{1}{N}\sum_{m=1}^{N}x_{m}^{(0)}(\theta)=\frac{1}{N}\sum_{m=1}^{N}\log p(y_{m}|\theta),
\end{equation}
from which after multiplication with $N$, adding $\log p(\theta)$
and taking exponentials, $p(\theta|\bm{y})$ can be determined at
each node.

\subsection{Uniformly Reweighted Belief Propagation Consensus\label{subsec:Belief-Propagation-Consensus}}

\begin{figure*}
\centering\includegraphics[width=0.75\paperwidth]{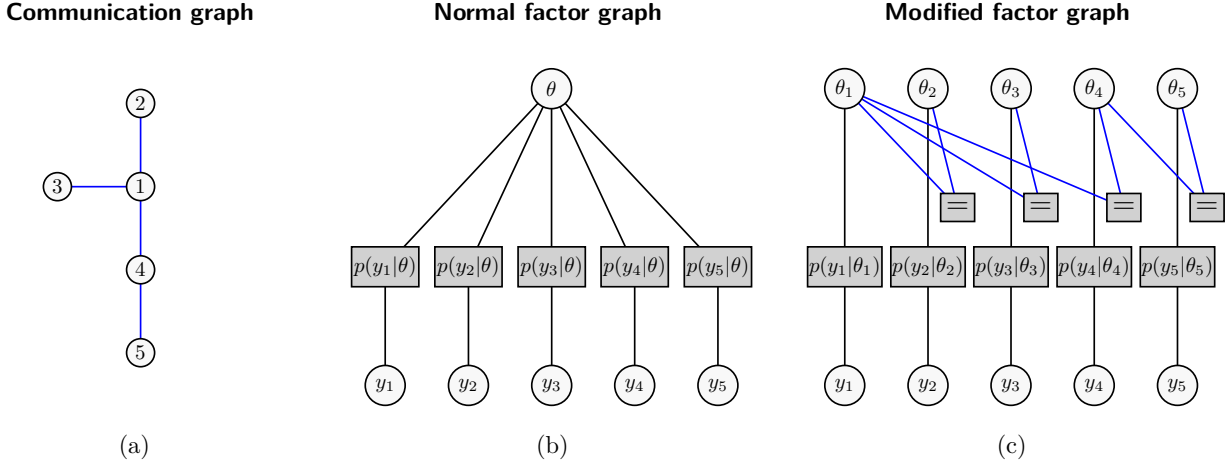}\caption{\label{fig:factorgraphs}This figure shows how the original factor
graph structure of the centralized problem (the left of the factor
graphs), with one $\theta$, is transformed into the factor graph
of its corresponding distributed problem (the graph on the right side),
as described in Section \ref{subsec:Belief-Propagation-Consensus},
with factor vertices as gray boxes and variable vertices as light
gray circles. Note that the connections between the equality factors
and the $\theta_{n}$'s are decided by the edge set $\mathcal{E}$.
The communication links are highlighted in blue to show how these
edges map to edges in the modified factor graph.}
\end{figure*}
 When expressing (\ref{eq:postfact}) as a factor graph, we obtain
a graph with a star topology, irrespective of the network graph. This
is shown in Fig.~\ref{fig:factorgraphs}: Fig.~\ref{fig:factorgraphs}\textendash (a)
shows a network graph and Fig.~\ref{fig:factorgraphs}\textendash (b)
shows the corresponding factor graph. Thus the structure of the factor
graph does not match the node graph $\mathcal{G}$. In order to obtain
a factor graph that matches the topology of the graph in Fig.~\ref{fig:factorgraphs}\textendash (a),
we introduce 
\begin{equation}
f(\theta_{1},\dots,\theta_{N})=\prod_{m=1}^{N}p(y_{m}|\theta_{m})\prod_{(m,n)\in\mathcal{E}}\mathbb{I}_{\{\theta_{m}=\theta_{n}\}},\label{eq:bpfact}
\end{equation}
which is shown in Fig.~\ref{fig:factorgraphs}\textendash (c). The
marginal of this function with respect to $\theta_{m}$ is given by
\begin{align}
f_{m}(\theta_{m}) & =\sum_{\sim\theta_{m}}f(\theta_{1},\dots,\theta_{N})=\prod_{l=1}^{N}p(y_{l}|\theta_{m}),
\end{align}
so that for any $\theta$, $f_{m}(\theta)=f_{n}(\theta)$, allowing
every node to determine the posterior by multiplying $f_{m}(\theta)$
with $p(\theta)$. The functions $f_{m}\left(\theta\right)$ can
be computed using message-passing algorithms, such as BP or URW-BP.
The initial belief of node $n$ in the log-domain is $x_{n}^{(0)}(\theta)=\log p(y_{n}|\theta)$.
By applying the URW-BP rules in the log-domain, we find (see Appendix
\ref{sec:bpcderivation}) 
\begin{align}
x_{n}^{(1)}(\theta) & =x_{n}^{(0)}(\theta)+\rho\sum_{m\in\mathcal{N}_{n}}x_{m}^{(0)}(\theta)\label{eq:bpcinit}\\
x_{n}^{(\ell)}(\theta) & =x_{n}^{(\ell-2)}(\theta)+\rho\sum_{m\in\mathcal{N}_{n}}(x_{m}^{(\ell-1)}(\theta)-x_{n}^{(\ell-2)}(\theta)),\label{eq:bpcupdate}
\end{align}
for $\ell>1$, where $\rho\in(0,1)$ is the reweighting parameter
of URW-BP (where standard BP corresponds to $\rho=1$). We call
the resulting algorithm uniformly reweighted belief propagation consensus
(URW-BPC), due to its linear update resembling a consensus algorithm.
 Defining the $2N\times2N$ URW-BPC matrix
\begin{equation}
\bm{P}_{\rho}=\left[\begin{array}{cc}
\rho\bm{A} & \bm{I}_{N}-\rho\bm{D}\\
\bm{I}_{N} & \bm{O}_{N}
\end{array}\right],
\end{equation}
the update rule in matrix form for URW-BPC for $\ell>1$ is
\begin{align}
\bm{x}^{(\ell)}(\theta) & =\left[\begin{array}{cc}
\bm{I}_{N} & \bm{O}_{N}\end{array}\right]\bm{P}_{\rho}\left[\begin{array}{c}
\bm{x}^{(\ell-1)}(\theta)\\
\bm{x}^{(\ell-2)}(\theta)
\end{array}\right]\\
 & =\left[\begin{array}{cc}
\bm{I}_{N} & \bm{O}_{N}\end{array}\right]\bm{P}_{\mathbf{\rho}}^{\ell-1}\left[\begin{array}{c}
\bm{x}^{(1)}(\theta)\\
\bm{x}^{(0)}(\theta)
\end{array}\right].
\end{align}
The convergence behavior depends on the power series of the update
matrix $\bm{P}_{\rho}$. 
\begin{rem}
We note that, due to (\ref{eq:bpcinit}), it holds that
\begin{equation}
\bm{x}^{(\ell)}(\theta)=\left[\begin{array}{cc}
\bm{I}_{N} & \bm{O}_{N}\end{array}\right]\bm{P}_{\rho}^{\ell-1}\left[\begin{array}{c}
\rho\bm{A}+\bm{I}_{N}\\
\bm{I}_{N}
\end{array}\right]\bm{x}^{(0)}(\theta),\label{eq:bpcfull}
\end{equation}
which can equivalently be expressed as 
\begin{equation}
\bm{x}^{(\ell)}(\theta)=\left[\begin{array}{cc}
\bm{I}_{N} & \bm{O}_{N}\end{array}\right]\bm{P}_{\rho}^{\ell-1}\left(\bm{P}_{\rho}+\bm{I}_{2N}\right)\left[\begin{array}{c}
\bm{x}^{(0)}(\theta)\\
\bm{0}
\end{array}\right].\label{eq:bpcinitsq}
\end{equation}
 
\end{rem}

\section{General Convergence Results for URW-BPC\label{sec:Eigenvalues-and-Jordan}}

Since URW-BPC results in an update rule that can be described in terms
of a matrix-vector multiplication, the convergence behavior depends
on how the power series $\bm{P}_{\rho}^{\ell}$ behaves as $\ell$
grows large, which will here be analyzed. First, we establish the
fact that $\lambda_{1}=1$ is an eigenvalue of any $\bm{P}_{\rho}$,
and give its corresponding right and left eigenvectors. 
\begin{prop}
\label{prop:eigv1}For any URW-BPC matrix $\bm{P}_{\rho}$ there is
one eigenvalue $\lambda_{1}=1$ with geometric multiplicity $1$.
Its corresponding right and left eigenvectors are $\bm{b}_{1}=\bm{1}$,
and $\bm{c}_{1}^{\mathsf{T}}=\left[\bm{1}^{\mathsf{T}},\bm{1}^{\mathsf{T}}-\rho\bm{1}^{\mathsf{T}}\bm{D}\right]$,
respectively.
\end{prop}
\begin{IEEEproof}
See Appendix \ref{subsec:Proof-of-PropositionA}.
\end{IEEEproof}
Hence, if all other eigenvalues are strictly less than 1, the convergence
of URW-BPC is guaranteed. In contrast to the $\bm{W}$ matrix in
belief consensus, the matrix $\bm{P}_{\rho}$ may not be diagonalizable.
Hence, we must consider two cases before providing general convergence
conditions. 

\subsection{Case 1: Diagonalizable $\bm{P}_{\rho}$}

If $\bm{P}_{\rho}$ is a diagonalizable matrix, then by eigendecomposition
we have that $\bm{P}_{\rho}^{\ell}=\bm{B}\bm{\Lambda}^{\ell}\bm{B}^{-1}$,
where the columns of $\bm{B}$ form an eigenbasis of $\bm{P}_{\rho}$
and $\bm{\Lambda}$ is a matrix with the eigenvalues of $\bm{P}_{\rho}$
on the diagonal. Let 
\begin{equation}
\bm{z}^{(0)}(\theta)=\left[\begin{array}{c}
\rho\bm{A}+\bm{I}_{N}\\
\bm{I}_{N}
\end{array}\right]\bm{x}^{(0)}(\theta),
\end{equation}
and express $\bm{z}^{(0)}(\theta)$ in the eigenbasis of $\bm{P}_{\rho}$
as $\bm{z}^{(0)}(\theta)=\bm{B}\bm{\alpha}$. Now we see that
\begin{equation}
\bm{z}^{(\ell)}(\theta)=\bm{P}_{\rho}^{\ell}\bm{z}^{(0)}(\theta)=\bm{B}\bm{\Lambda}^{\ell}\bm{B}^{-1}\bm{B}\bm{\alpha}=\bm{B}\bm{\Lambda}^{\ell}\bm{\alpha}.
\end{equation}
We can also express this as
\begin{equation}
\bm{z}^{(\ell)}(\theta)=\sum_{i=1}^{2N}\lambda_{i}^{\ell}\bm{b}_{i}\alpha_{i},
\end{equation}
where $\lambda_{i}$ is the $i$th eigenvalue of $\bm{P}_{\rho}$,
$\bm{b}_{i}$ is the $i$th eigenvector of $\bm{P}_{\rho}$ (and the
$i$th column of $\bm{B}$), and $\alpha_{i}$ is the $i$th element
of $\bm{\alpha}$. Since according to Proposition \ref{prop:eigv1},
$\lambda_{1}=1$, so that 
\begin{align}
\bm{z}^{(\ell)}(\theta) & =\bm{b}_{1}\alpha_{1}+\sum_{i=2}^{2N}\lambda_{i}^{\ell}\bm{b}_{i}\alpha_{i}\label{eq:lincombeig}\\
 & =\bm{b}_{1}\alpha_{1}+\varepsilon.
\end{align}
Later, in Proposition \ref{prop:eigv2-1} we will establish that $\alpha_{1}$
is the sought value, therefore we consider $\varepsilon$ to be an
error term.

\textcolor{violet}{}

\subsection{Case 2. Nondiagonalizable $\bm{P}_{\rho}$}

If $\bm{P}_{\rho}$ is not diagonalizable, it can be decomposed in
its Jordan normal form. Then, $\bm{P}_{\rho}=\bm{B}\bm{J}\bm{B}^{-1}$,
where the columns of $\bm{B}$ are the generalized eigenvectors of
$\bm{P}_{\rho}$ forming a Jordan basis, and $\bm{J}$ is a Jordan
matrix, which is a block diagonal matrix with $M<2N$ Jordan blocks
on its diagonal, i.e.,
\begin{equation}
\bm{J}=\left[\begin{array}{ccc}
\bm{J}_{1} & 0 & 0\\
0 & \ddots & 0\\
0 & 0 & \bm{J}_{M}
\end{array}\right].
\end{equation}
Each Jordan block corresponds to a certain eigenvalue and its generalized
eigenvectors. For example, if the eigenvalue $\lambda_{m}$ has three
generalized eigenvectors, $\bm{b}_{m,1}$, $\bm{b}_{m,2}$ and $\bm{b}_{m,3}$,
then 
\[
\bm{J}_{m}=\left[\begin{array}{ccc}
\lambda_{m} & 1 & 0\\
0 & \lambda_{m} & 1\\
0 & 0 & \lambda_{m}
\end{array}\right].
\]
 Note that if $\bm{P}_{\rho}$ is diagonalizable, its Jordan normal
form is equal to its eigendecomposition. By expressing $\bm{z}^{(0)}(\theta)=\bm{B}\bm{\alpha}$
in the Jordan basis of $\bm{P}_{\rho}$ and decomposing $\bm{P}_{\rho}$
in Jordan normal form, we can write
\begin{equation}
\bm{z}^{(\ell)}(\theta)=\bm{P}_{\rho}^{\ell}\bm{z}^{(0)}(\theta)=\bm{B}\bm{J}^{\ell}\bm{\alpha},
\end{equation}
which we can also express as
\begin{equation}
\bm{z}^{(\ell)}(\theta)=\sum_{m=1}^{M}\sum_{j=1}^{r_{m}}\left(\sum_{i=0}^{\min(\ell,r_{m}-j)}\left(\begin{array}{c}
\ell\\
i
\end{array}\right)\lambda_{m}^{\ell-i}\right)\bm{b}_{m,j}\alpha_{m,j},\label{eq:lincombjordan}
\end{equation}
where $r_{m}$ is the size of the $m$th Jordan block, $\bm{b}_{m,j}$
is the $j$th generalized eigenvector of $\lambda_{m}$, and $\alpha_{m,j}$
the corresponding entry in $\bm{\alpha}$. With $\lambda_{1}=1$ and
denoting $\bm{b}_{1,1}\alpha_{1,1}$ by $\bm{b}_{1}\alpha_{1}$, we
can break the sum into three parts
\begin{align}
\bm{z}^{(\ell)}(\theta) & =\bm{b}_{1}\alpha_{1}+\sum_{j=2}^{r_{1}}\sum_{i=0}^{\min(\ell,r_{1}-j)}\left(\begin{array}{c}
\ell\\
i
\end{array}\right)\bm{b}_{1,j}\alpha_{1,j}\nonumber \\
 & +\sum_{m=2}^{M}\sum_{j=1}^{r_{m}}\left(\sum_{i=0}^{\min(\ell,r_{m}-j)}\left(\begin{array}{c}
\ell\\
i
\end{array}\right)\lambda_{m}^{\ell-i}\right)\bm{b}_{m,j}\alpha_{m,j}\label{eq:jordanlincomb}\\
 & =\bm{b}_{1}\alpha_{1}+\tilde{\varepsilon}+\varepsilon.\label{eq:jordanerror}
\end{align}
Following the reasoning of the case with a diagonalizable $\bm{P}_{\rho}$,
any quantity that is not $\alpha_{1}$ is considered an error term.
For the nondiagonalizable case, we split it up into $\varepsilon$
and $\tilde{\varepsilon}$, since these two terms behave fundamentally
different with respect to the eigenvalues of $\bm{P}_{\rho}$.

\subsection{General Convergence Conditions\label{subsec:Convergence-Section}}

We are now able to provide insights into $\alpha_{1}$ as well as
the error terms $\tilde{\varepsilon}$ and $\varepsilon$. 
\begin{prop}
\label{prop:eigv2-1}The quantity $\alpha_{1}^{(\ell)}=\bm{c}_{1}^{\mathsf{T}}\bm{z}^{(\ell)}(\theta)/(\bm{c}_{1}^{\mathsf{T}}\bm{b}_{1})$
is preserved by the URW-BPC algorithm at each iteration $\ell$. If
URW-BPC converges, then the consensus value is the preserved quantity,
and it is equal to
\begin{equation}
\alpha_{1}=\frac{2}{\bm{c}_{1}^{\mathsf{T}}\bm{b}_{1}}\bm{1}^{\mathsf{T}}\bm{x}^{(0)}\left(\theta\right).\label{eq:consensusvalue}
\end{equation}
\end{prop}
\begin{IEEEproof}
See Appendix \ref{subsec:Proof-of-PropositionB}. 
\end{IEEEproof}
Note according to Proposition \ref{prop:eigv1} 
\begin{equation}
\bm{c}_{1}^{\mathsf{T}}\bm{b}_{1}=2N-\rho\,\mathrm{trace}(\bm{D}).\label{eq:innerprod}
\end{equation}

Hence, what remains is to establish sufficient conditions for URW-BPC
to converge and then to establish the corresponding convergence rate.
We note the following:
\begin{enumerate}
\item When $\bm{P}_{\rho}$ has an eigenvalue $\lambda=-1$ \textcolor{black}{with
equal geometric and algebraic multiplicities,} the corresponding value
$\alpha_{i}$ in (\ref{eq:lincombeig}) or $\alpha_{m,j}$ in (\ref{eq:lincombjordan})
is zero, since the eigenvector of $\lambda=-1$ is in the null space
of $\bm{P}_{\rho}+\bm{I}_{2N}$ and is thus canceled out by the initialization
(\ref{eq:bpcinitsq}). 
\item When $\bm{P}_{\rho}$ is diagonalizable, there is only one eigenvalue
$\lambda_{1}=1$. If all other eigenvalues are strictly inside the
unit circle, or equal to $-1$ \textcolor{black}{with equal geometric
and algebraic multiplicities,} then $\varepsilon\to0$ and convergence
of URW-BPC (\ref{eq:bpcfull}) is guaranteed to (\ref{eq:consensusvalue})
by Proposition \ref{prop:eigv2-1}.
\item When $\bm{P}_{\rho}$ is not diagonalizable, if $\lambda_{1}$ has
a Jordan block of size $1\times1$ and all other eigenvalues are strictly
inside the unit circle, or equal to $-1$ \textcolor{black}{with equal
geometric and algebraic multiplicities,} then $\tilde{\varepsilon}=0$,
$\varepsilon\to0$, and convergence of URW-BPC (\ref{eq:bpcfull})
is guaranteed to (\ref{eq:consensusvalue}) by Proposition \ref{prop:eigv2-1}.
\end{enumerate}
\textcolor{blue}{}\textcolor{black}{}

Finally, the convergence rate is defined as
\begin{equation}
r\left(\bm{P}_{\rho}\right)=\sup_{\bm{x}^{(0)}(\theta)\neq c\bm{1}}\lim_{\ell\to\infty}\left(\frac{\left\Vert \bm{x}^{(\ell)}(\theta)-c\bm{1}\right\Vert _{2}}{\left\Vert \bm{x}^{(0)}(\theta)-c\bm{1}\right\Vert _{2}}\right)^{1/\ell},
\end{equation}
provided that the algorithm is convergent, and at least one eigenvalue
strictly inside the unit circle is nonzero. For such cases, we consider
the eigenvalues of $\bm{P}_{\rho}$ to be sorted such that $\left|\lambda_{1}\right|\ge\left|\lambda_{2}\right|\ge\dots\ge\left|\lambda_{2N}\right|$.
The convergence rate is determined by $|\tilde{\lambda}|$ where $\tilde{\lambda}=\max_{i}|\lambda_{i}|$
for $i$ such that $|\lambda_{i}|<1$, such that a smaller $|\tilde{\lambda}|$
gives a faster convergence.

\section{Convergence on Specific Graph Types\label{sec:Convergence-on-Tree-graphs}}

In this section, we analyze the convergence properties of URW-BPC
for three specific types of graphs. We first consider tree-graphs,
recovering the well-known finite-time BP convergence result via the
formulation (\ref{eq:bpcinitsq}). Then, we consider the regular graphs,
for which BP is generally not guaranteed to converge. Finally, we
consider general connected graphs, for which we can build on the results
from regular graphs. 

\subsection{Tree Graphs}

For trees, the following proposition establishes the possible eigenvalues
of $\bm{P}_{1}$. 
\begin{prop}
\label{lem:tree-eigenvalues}The URW-BPC matrix $\bm{P}_{1}$ of any
tree-graph has three distinct eigenvalues: $\lambda_{1}=1$, $\lambda_{2}=-1$,
and $\lambda_{i}=0$ for $i=3,\dots,2N$.
\end{prop}
\begin{IEEEproof}
See Appendix \ref{subsec:Proof-of-LemmaTREE}.
\end{IEEEproof}
We then immediately find the following well-known results for
trees, that BP converges in a finite number of iterations. 
\begin{thm}
\label{thm:tree-g-true}If $\mathcal{G}$ is a tree graph of $N$
nodes, then URW-BPC with $\bm{P}_{1}$ converges to consensus after
at most $2N-3$ iterations. Moreover, with the initialization as in
(\ref{eq:bpcfull}), the consensus value after $\kappa$ iterations
($\kappa$ such that consensus is reached) is
\begin{equation}
\bm{x}^{(\kappa)}(\theta)=\sum_{m=1}^{N}x_{m}^{(0)}(\theta)\bm{1}.\label{eq:treeconsensusvalue}
\end{equation}

\textcolor{blue}{}
\end{thm}
\begin{IEEEproof}
Due to Proposition \ref{lem:tree-eigenvalues}, $\bm{P}_{1}$ has
$2N-2$ eigenvalues $\lambda=0$. Hence, the largest possible size
of its corresponding Jordan block, denoted by $\bm{J}_{0}$, is $2N-2$.
Since $\bm{J}_{0}^{2N-2}=\bm{O}$, the error contribution from the
eigenvalues equal to zero is zero after at most $2N-3$ iterations.
Furthermore, applying the results from Propositions \ref{prop:eigv1}
and \ref{prop:eigv2-1}, and using the fact that the sum of the degrees
$\mathrm{trace}(\bm{D})=2N-2$ for undirected tree-graphs in (\ref{eq:innerprod}),
the consensus value $\alpha_{1}$ is given by (\ref{eq:treeconsensusvalue}). 
\end{IEEEproof}

\subsection{Regular Graphs\label{subsec:Regular-Graphs}}

In order to understand when URW-BPC converges, we first show how to
choose the weighting parameter $\rho$ in order to guarantee convergence.
Then we proceed to optimize $\rho$ for a given graph $\mathcal{G}$
such that the magnitude of the largest eigenvalue inside the unit
circle, $|\tilde{\lambda}|$, is minimized. We recall that for $k$-regular
graphs, the largest eigenvalue of the adjacency matrix is $\mu_{1}=k$
for non-bipartite graphs, while for bipartite graphs, eigenvalues
come in symmetric pairs, so that both $\mu_{1}=k$ and $\mu_{2}=-k$
are eigenvalues \cite[Prop.2.3]{lovasz2007eigenvalues}. 

\subsubsection{Convergence}

To find for which $\rho$ URW-BPC converges on $k$-regular graphs,
we first show how the eigenvalues of $\bm{P}_{\rho}$ and $\bm{A}$
are connected in terms of magnitudes. \textcolor{black}{Note, that
the eigenvalues of $\bm{P}_{\rho}$ and $\bm{A}$ are sorted such
that $|\lambda_{1}|\ge|\lambda_{2}|\ge\dots\ge|\lambda_{2N}|$ and
$|\mu_{1}|\ge|\mu_{2}|\ge\dots\ge|\mu_{N}|$.}
\begin{lem}
\label{lem:secondlargest}Let $\rho\in(0,1]$. Then the eigenvalue
$\lambda_{i}$ of $\bm{P}_{\rho}$, \textcolor{black}{$\lambda_{i}\neq0$,
can be expressed in terms of $\mu_{i}$,} and its magnitude is
\begin{equation}
\left|\lambda_{i}\right|=\frac{1}{2}\left|\mu_{i}\rho+\sqrt{\mu_{i}^{2}\rho^{2}-4k\rho+4}\right|.\label{eq:EVrelation}
\end{equation}
\end{lem}
\begin{IEEEproof}
See Appendix \ref{subsec:secondlargets-proof}.
\end{IEEEproof}
\textcolor{black}{Note that for $\lambda_{i}=0$ with eigenvector
$[\bm{v}^{\mathsf{T}},\bm{w}^{\mathsf{T}}]^{\mathsf{T}}$, we have
that
\begin{align}
\rho\bm{A}\bm{v}+\bm{w}-\rho\bm{D}\bm{w} & =0\\
\bm{v} & =0.
\end{align}
We conclude that for $\lambda_{i}=0$, we must have that $k\rho=1$.
Note however, that this does not mean that all eigenvalues are equal
to 0 for $\rho=1/k$.}

Now, since $\mu_{1}=k$, the magnitude of $\lambda_{1}$ of $\bm{P}_{\rho}$
is either $1$ or $\left|\rho k-1\right|$. Thus, we can prove the
following result regarding the convergence conditions of URW-BPC.
\begin{thm}
\label{lem:kconvergence}For any $k$-regular graph, URW-BPC is convergent
if and only if $\rho\in(0,2/k)$, and the asymptotic consensus value
is
\begin{equation}
\lim_{\ell\to\infty}\bm{x}^{(\ell)}\left(\theta\right)=\frac{1}{N(1-\rho k/2)}\sum_{m=1}^{N}x_{m}^{(0)}\left(\theta\right)\bm{1}.
\end{equation}
\end{thm}
\begin{IEEEproof}
See Appendix \ref{subsec:kconvergence-proof}.
\end{IEEEproof}
This result provides the interval for $\rho$ within which we can
guarantee convergence, and to which value the algorithm converges. 

\textcolor{blue}{}

\subsubsection{Optimizing the Convergence Rate}

In order to maximize convergence rate, we show which $\rho$ minimizes
the largest eigenvalue within the unit circle, denoted by $|\tilde{\lambda}|<1$.
\begin{thm}
\label{thm:optrho}The choice of $\rho$ that minimizes $|\tilde{\lambda}|$
is
\begin{equation}
\rho_{\mathrm{opt}}=\frac{2}{\tilde{\mu}{}^{2}}\left(k-\sqrt{k^{2}-\tilde{\mu}^{2}}\right),\label{eq:rhoopt}
\end{equation}
where $\tilde{\mu}=\max_{i}\left|\mu_{i}\right|$ for $i$ such that
$\left|\mu_{i}\right|<k$. The magnitude of the second largest eigenvalue
of $\bm{P}_{\rho_{\mathrm{opt}}}$ is
\begin{align}
|\tilde{\lambda}| & =\left|\frac{1}{\tilde{\mu}}\left(k-\sqrt{k^{2}-\tilde{\mu}^{2}}\right)\right|.\label{eq:l2opt}
\end{align}
\end{thm}
\begin{IEEEproof}
See Appendix \ref{subsec:Proof-of-TheoremCR}.
\end{IEEEproof}
\begin{rem}
For any $k$-regular \emph{non-bipartite} graph $\mathcal{G}$, $\tilde{\mu}=\mu_{2}$.
However, for a $k$-regular \emph{bipartite} graph we have that $\mu_{\mathrm{2}}=-\mu_{1}=-k$.
Hence, choosing $\rho_{\mathrm{opt}}$ with $\mu_{2}$ instead of
$\tilde{\mu}$ in this case would yield $\rho_{\mathrm{opt}}=2/k$,
which in turn gives (see (\ref{eq:EVrelation})), $\left|\lambda_{i}\right|^{2}=1/k^{2}\left|\mu\pm\sqrt{\mu^{2}-k^{2}}\right|^{2}=1$,
for every eigenvalue $\lambda_{i}$ of $\bm{P}_{\rho}$. Hence, the
optimal reweighting for $k$-regular bipartite graphs is achieved
with $\tilde{\mu}=\mu_{3}$. Another consequence of $\mu_{2}=-k$
is that there is always an eigenvalue $\lambda=-1$ for bipartite
graphs. \textcolor{green}{}This remark also applies to tree-graphs,
which is a class of bipartite graphs, where $\bm{P}_{1}$ of a tree-graph
has an eigenvalue $\lambda=-1$. However, the component associated
with this eigenvalue is irrelevant, as it is removed by the initialization
procedure. \textcolor{black}{This relies on the following result.}
\end{rem}
\begin{prop}
\textcolor{blue}{\label{prop:bipartiteeig}}\textcolor{black}{For
a URW-BPC matrix $\bm{P}_{\rho}$, the algebraic and geometric multiplicites
of $\lambda=\pm1$ are equal.}
\end{prop}
\begin{IEEEproof}
\textcolor{black}{See Appendix \ref{subsec:bipartiteeig-proof}.}
\end{IEEEproof}

\subsubsection{Limit Results for $k$-regular Graphs\label{subsec:Limit-Results-for}}

Due to their structure, the eigenvalue distribution of $\bm{A}$
for large $k$-regular graphs is given by \cite{mckay1981expected}
(with $\mathcal{G}$ satisfying certain properties regarding the number
of cycles in the graph, for details see \cite{mckay1981expected})
\begin{equation}
f(\mu)=\begin{cases}
\frac{k\left(4\left(k-1\right)-\mu^{2}\right)^{1/2}}{2\pi\left(k^{2}-\mu^{2}\right)}, & \left|\mu\right|\le2\sqrt{k-1}\\
0, & \left|\mu\right|>2\sqrt{k-1}.
\end{cases}
\end{equation}
This means that $\lim_{N\to\infty}\left|\mu_{2}\right|=2\sqrt{k-1}$,
and thus
\begin{equation}
\lim_{N\to\infty}\rho_{\mathrm{opt}}=\frac{1}{2(k-1)}\left(k-\sqrt{k^{2}-4(k-1)}\right),
\end{equation}
so that the second largest eigenvalue of $\bm{P}_{\rho}$ for non-bipartite
graphs tends to
\begin{equation}
\lim_{N\to\infty}\left|\lambda_{2,\mathrm{BPC}}\right|=\frac{1}{2\sqrt{k-1}}\left(k-\sqrt{k^{2}-4\left(k-1\right)}\right).
\end{equation}
For the belief consensus, $\left|\lambda_{2,\mathrm{Metr}}\right|$
of $\bm{W}=\bm{I}_{N}-\xi\bm{L}$ tends to
\begin{equation}
\lim_{N\to\infty}\left|\lambda_{2,\mathrm{Metr}}\right|=\frac{1+2\sqrt{k-1}}{k+1}>\lim_{N\to\infty}\left|\lambda_{2,\mathrm{BPC}}\right|,
\end{equation}
so that BPC always converges faster than belief consensus on large
$k$-regular graph.\textcolor{red}{{} }

\subsection{General Graphs\label{subsec:selfloops}}

For general graphs, it is not obvious how to render BPC convergent.
A possible approach is to determine a spanning tree of the network
graph and then running BPC with finite-time convergence \cite{savic2010indoor}.
However, we can also build on the results from regular graphs. We
outline two procedures to convert a general graph to a regular graph. 
\begin{enumerate}
\item \emph{Edge addition: }The simplest way to make a graph into a $k$-regular
graph, is to first determine the maximum node degree $d_{\mathrm{max}}$
(this can be done through max-consensus). Then a node $i$ with degree
$d_{i}$ adds $d_{\mathrm{max}}-d_{i}$ self-loops. Then BPC with
$\rho_{\mathrm{opt}}$ set based on $k=d_{\mathrm{max}}$ and $\tilde{\mu}$
of the new $\bm{A}$, is applied. 
\item \emph{Edge deletion:} A more complex way to create a $k$-regular
graph is by selectively deleting edges from those nodes with maximum
degree, while maintaining connectivity. This procedure can be applied
until a certain minimal value for $d_{\mathrm{max}}$ is attained. 
\end{enumerate}

\section{Numerical Results and Discussion\label{sec:Results-and-Discussion}}

\subsection{Simulation Parameters}

We present numerical results comparing the URW-BPC algorithms with
Metropolis weighted belief consensus. The simulations were performed
with the number of nodes $N=100$, with a fixed $\mathcal{G}$ for
tree-graphs and random $\mathcal{G}$ for $k$-regular graphs. The
node degree for the $k$-regular graphs was fixed to $k=4$. The elements
of the initial data $\bm{x}^{(0)}(\theta)$ were generated according
to a standard normal distribution. We calculated the averaged (over
instances of $\bm{x}^{(0)}(\theta)$ for tree-graphs, and over both
$\bm{x}^{(0)}(\theta)$ and $\mathcal{G}$ for $k$-regular graphs)
and normalized mean squared error (MSE), with the MSE being normalized
with respect to the initial consensus error. The simulations were
performed over 100 Monte Carlo runs. \textcolor{magenta}{}

\subsection{Results for Tree-graphs}

\begin{figure}
\includegraphics[width=1\columnwidth]{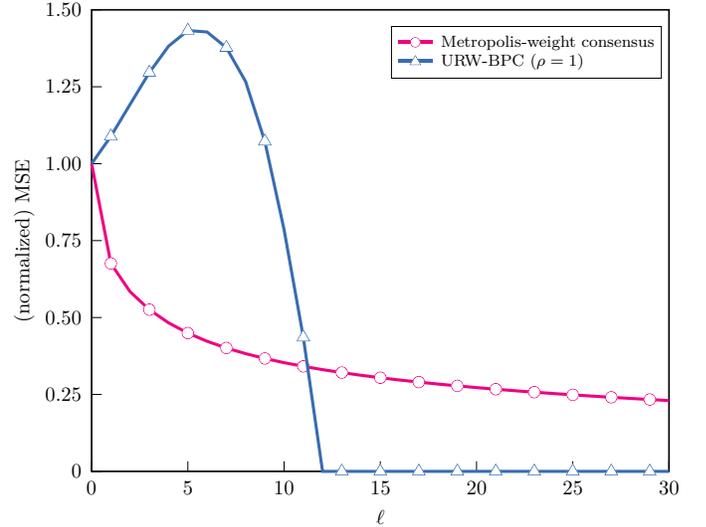}

\caption{\label{fig:tree-sim}The normalized MSE of URW-BPC vs. BC with Metropolis-type
weights over a random tree-graph, with diameter $12$ and $N=100$.
The error is averaged over $100$ instances of random initial data
$\bm{x}^{(0)}(\theta)$, but with a fixed graph $\mathcal{G}$.}

\end{figure}
 The simulated error of URW-BPC on a tree-graph is shown in Fig.~\ref{fig:tree-sim}.
We observe that the algorithm indeed reaches consensus in a finite
number of steps. However, before reaching consensus, the error of
URW-BPC behaves differently from that of the other consensus algorithm.
The increasing error we see can be explained by $\varepsilon_{1}$
in (\ref{eq:jordanerror}). It takes a few iterations for the Jordan
blocks of the eigenvalues $\lambda_{i}=0$ to become zero, and until
they do, the error they contribute with increases as $k$ increases.

\subsection{Results for $k$-regular Graphs\label{subsec:k-reg results}}

\textcolor{green}{}
\begin{figure}
\includegraphics[width=1\columnwidth]{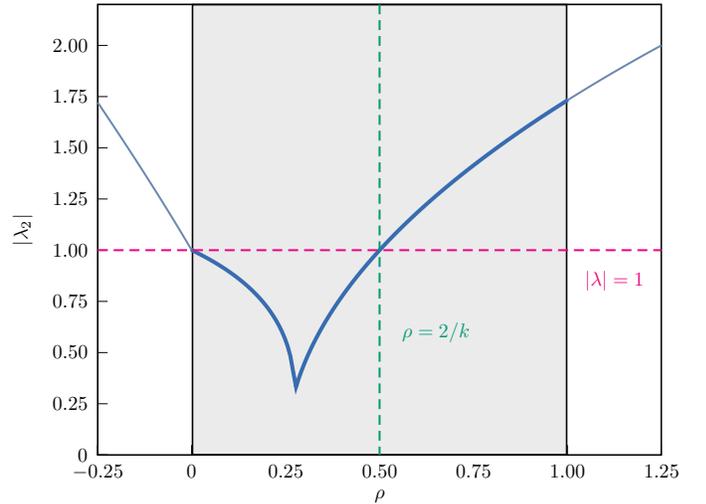}

\caption{\label{fig:The-magnitude-of}The magnitude of the eigenvalue $\lambda_{2}$
of $\bm{P}_{\rho}$ generated from the eigenvalue $\mu_{2}$ of $\bm{A}$
plotted as a function of the weighting parameter $\rho$, for the
$k$-regular small-world graph with $k=4$ and $N=10$ described in
the example in Section \ref{subsec:k-reg results}.}
\end{figure}
\begin{figure}
\includegraphics[width=1\columnwidth]{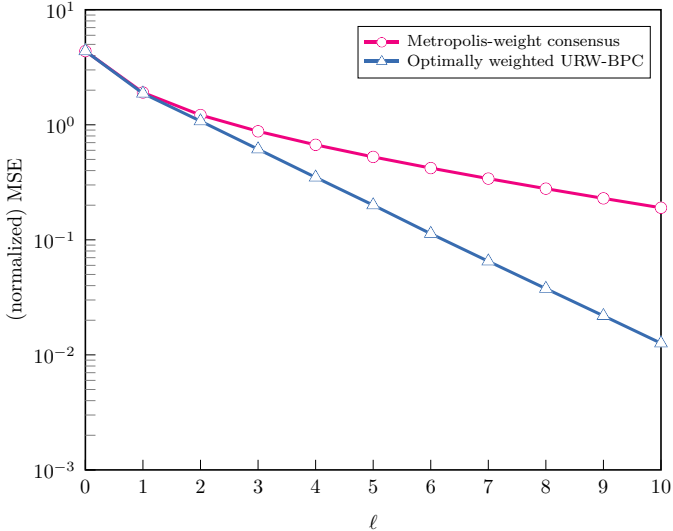}\caption{\label{fig:reg-sim}The normalized MSE of URW-BPC with $\rho=\rho_{\mathrm{opt}}$
vs. BC with Metropolis-type weights over random regular graphs with
$N=100$ and $k=4$, plotted in log scale as a function of iterations.
The error is averaged over $100$ instances of random initial data
$\bm{x}^{(0)}(\theta)$, and graphs $\mathcal{G}$.}
\end{figure}
\begin{figure}
\includegraphics[width=1\columnwidth]{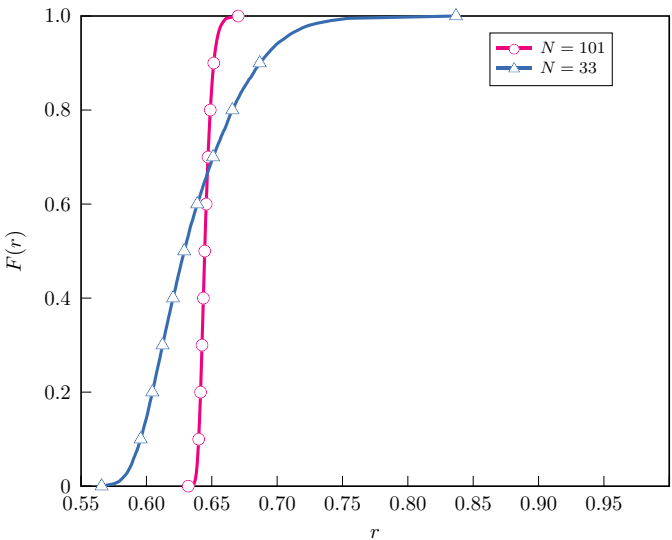}

\caption{\label{fig:Empirical-cdf's-of}Empirical cdf's of $r=\left|\lambda_{2,\mathrm{BPC}}\right|/\left|\lambda_{2,\mathrm{Metr}}\right|$
for two types of $k$-regular graphs with $N=101$ (red) and $N=33$
(blue), and $k=4$. The empirical cdf is from $1000$ instances of
random graphs.}
\end{figure}
To illustrate the benefit of URW-BPC on $k$-regular graphs, we provide
the following example of URW-BPC on a so called small-world graph
\cite{farkas2001spectra}. Let $\mathcal{G}$ be of the type detailed
in \cite[Appendix A]{farkas2001spectra}. For this type of graph,
there exist closed-form expressions for the eigenvalues of $\bm{A}$.
In particular, if we let $\mathcal{G}$ be such a graph with $N=10$
and $k=4$, we have that $\mu_{2}\approx2.23$. Hence, using (\ref{eq:rhoopt})
to calculate the optimal $\rho$, we get from (\ref{eq:l2opt}) that
$\left|\lambda_{2}\right|\approx0.31$. On the other hand, using BC
on this graph with step-size $\varepsilon=0.25$ yields $\left|\lambda_{2}\right|\approx0.56$.

In Fig.~\ref{fig:reg-sim} we show how the average error of optimally
weighted URW-BPC and belief consensus compare. The error is averaged
over instances of the graph $\mathcal{G}$ as well as the initial
data $\bm{x}^{(0)}(\theta)$. Clearly, URW-BPC outperforms belief
consensus in terms of convergence rate. 

In Fig. \ref{fig:Empirical-cdf's-of} we compare the magnitude of
$\lambda_{2}$ for the two consensus algorithms by taking the ratio
$r=\left|\lambda_{2,\mathrm{BPC}}\right|/\left|\lambda_{2,\mathrm{Metr}}\right|$,
plotting the empirical cumulative distribution function (cdf) over
10,000 Monte Carlo runs. Clearly, URW-BPC always outperforms belief
consensus, since the ratio stays well below one. Moreover, as we increase
the network size, we see that the ratio converges to the specific
limit value discussed in Section \ref{subsec:Limit-Results-for}.

\subsection{Results for a General Graph}

\begin{figure}
\centering\includegraphics[width=0.5\columnwidth]{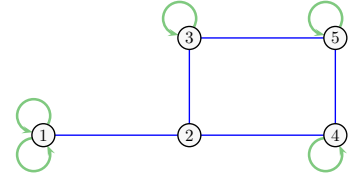}

\caption{\label{fig:selfloops_graph}Example of nonregular graph made $k$-regular
by adding self loops. In this case $k=3$. The original communication
graph is indicated by the blue edges, while the added self-loops are
indicated in green.}
\end{figure}
\begin{figure}
\includegraphics[width=1\columnwidth]{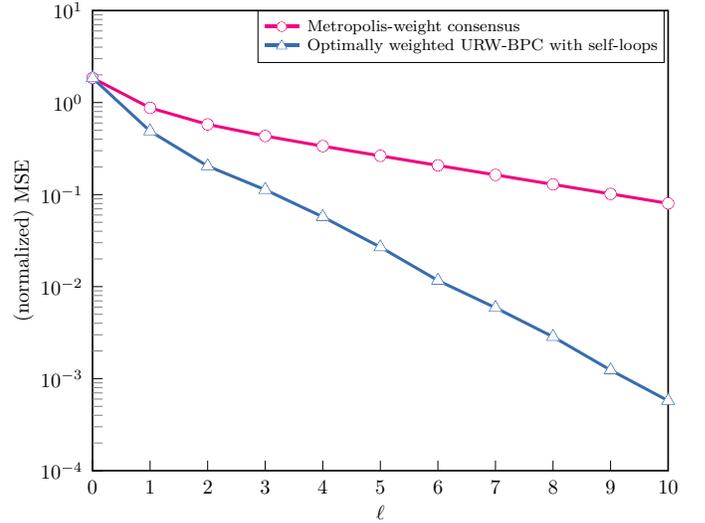}

\caption{\label{fig:selfloops_result}The normalized MSE of URW-BPC with $\rho=\rho_{\mathrm{opt}}$
with added self-loops to make the graph $k$-regular vs. BC with Metropolis-type
weights, with $N=100$ and $k=3$, plotted in log scale as a function
of iterations. The error is averaged over $100$ instances of random
initial data $\bm{x}^{(0)}(\theta)$.}
\end{figure}
To illustrate the method for general graphs discussed in Section \ref{subsec:selfloops},
we perform numerical simulations for the graph shown in Fig.~\ref{fig:selfloops_graph},\textbf{
}both without self-loops and Metropolis weight consensus, and with
added self-loops and optimally weighted (according to the result of
Theorem \ref{thm:optrho}) URW-BPC. The results are shown in Fig.~\ref{fig:selfloops_result}.
We see that the strategy of adding self-loops and running URW-BPC
on the resulting $k=3$ regular graph indeed works well, and asymptotically
outperforms belief consensus on the original graph.

\section{Conclusion\label{sec:Conclusion}}

We studied the uniformly reweighted belief propagation algorithm for
distributed likelihood fusion, which was described by a factor graph
with strong interactions, generally considered a challenging case
for belief propagation. The belief propagation consensus algorithm
resulted in a linear update rule much like a consensus algorithm with
memory. By eigenvalue analysis we were able to prove a collection
of results on several types of graphs: (i) we recovered the classical
finite-time convergence of belief propagation on tree graphs for the
likelihood fusion problem; (ii) we provided conditions on the reweighting
parameter necessary and sufficient for convergence, and (iii) we found
an analytical expression for the reweighting parameter optimizing
convergence rate, on $k$-regular graphs, and on general graphs artificially
transformed into $k$-regular graphs by adding self-loops or removing
edges. Based on both numerical results, and eigenvalue limits on large
$k$-regular graphs, belief propagation consensus outperformed consensus
with Metropolis-type weights. Open issues include analytically comparing
the performance of belief propagation consensus to other algorithms
for distributed likelihood fusion, and to investigate how it compares
to consensus algorithms with memory.

\appendices{}

\section{Derivation of the URW-BPC algorithm\label{sec:bpcderivation}}

According to the message-passing equations of the uniformly reweighted
BP \cite{wymeersch2012uniformly}, we can write the marginal belief
of some variable $\theta_{n}$ of node $n$ at iteration $\ell$ as
\begin{equation}
b_{n}^{(\ell)}(\theta_{n})\propto p(y_{n}|\theta_{n})\prod_{m\in\mathcal{N}_{n}}\left(\mu_{m\rightarrow n}^{(\ell)}(\theta_{n})\right)^{\rho},\label{eq:marginal}
\end{equation}
for $\rho\in(0,1]$, where the message from node $m$ to node $n$
at iteration $\ell$ is computed by
\begin{align}
\mu_{m\rightarrow n}^{(\ell)}(\theta_{n}) & \propto\sum_{\theta_{m}}\mathbb{I}_{\{\theta_{m}=\theta_{n}\}}p(y_{m}|\theta_{m})\frac{\prod_{u\in\mathcal{N}_{m}\backslash n}\left(\mu_{u\rightarrow m}^{(\ell-1)}(\theta_{m})\right)^{\rho}}{\left(\mu_{n\rightarrow m}^{(\ell-1)}(\theta_{m})\right)^{1-\rho}}\\
 & =\sum_{\theta_{m}}\mathbb{I}_{\{\theta_{m}=\theta_{n}\}}\frac{b_{m}^{(\ell-1)}(\theta_{m})}{\left(\mu_{n\rightarrow m}^{(\ell-1)}(\theta_{m})\right)^{1-\rho}\left(\mu_{n\rightarrow m}^{(\ell-1)}(\theta_{m})\right)^{\rho}}\\
 & =\frac{b_{m}^{(\ell-1)}(\theta_{n})}{\mu_{n\rightarrow m}^{(\ell-1)}(\theta_{n})}.\label{eq:message}
\end{align}
We note that $\theta_{n}=\theta_{m}=\theta$, and plug (\ref{eq:message})
into (\ref{eq:marginal})
\begin{align}
b_{n}^{(\ell)}(\theta) & \propto p(y_{n}|\theta)\prod_{m\in\mathcal{N}_{n}}\left(\frac{b_{m}^{(\ell-1)}(\theta)}{\mu_{n\rightarrow m}^{(\ell-1)}(\theta)}\right)^{\rho}\label{eq:marg}\\
 & =p(y_{n}|\theta)\prod_{m\in\mathcal{N}_{n}}\left(\frac{b_{m}^{(\ell-1)}(\theta)}{b_{n}^{(\ell-2)}(\theta)}\mu_{m\rightarrow n}^{(\ell-2)}(\theta)\right)^{\rho}\\
 & =b_{n}^{(\ell-2)}(\theta)\prod_{m\in\mathcal{N}_{n}}\left(\frac{b_{m}^{(\ell-1)}(\theta)}{b_{n}^{(\ell-2)}(\theta)}\right)^{\rho}.
\end{align}
For the initial values of the marginals, we assume that $\mu_{m\rightarrow n}^{(0)}(\theta)=1$
for all nodes $n$, all $m\in\mathcal{N}_{n}$. Hence, by (\ref{eq:marginal})
we have that $b_{n}^{(0)}(\theta)=p(y_{n}|\theta)$. Now we can compute
the marginals at iteration $\ell=1$ by using (\ref{eq:marg}), which
gives $b_{n}^{(2)}(\theta)=p(y_{n}|\theta)\prod_{m\in\mathcal{N}_{n}}\left(p(y_{m}|\theta)\right)^{\rho}$.

\section{Proofs of Propositions}

\subsection{\label{subsec:Proof-of-PropositionA}Proof of Proposition \ref{prop:eigv1}}
\begin{IEEEproof}
Let $\bm{b}_{1}=[\bm{v}^{\mathsf{T}},\bm{w}^{\mathsf{T}}]^{\mathsf{T}}$
be a right eigenvector corresponding to the eigenvalue $\lambda_{1}=1$.
Then, it holds that
\begin{align}
\bm{v} & =\rho\bm{A}\bm{v}+(\bm{I}_{N}-\rho\bm{D})\bm{w}\\
\bm{w} & =\bm{v},\label{eq:eigv1-1}
\end{align}
which boils down to $\bm{L}\bm{v}=\bm{0}$, where $\bm{L}=\bm{D}-\bm{A}$
is the graph Laplacian. The graph Laplacian $\bm{L}$ of a connected
graph has an eigenvalue $\nu=0$ with algebraic multiplicity equal
to 1, and its right eigenvector is $\bm{v}=\bm{1}$. Thus, $\lambda_{1}=1$
has geometric multiplicity equal to 1. Since (\ref{eq:eigv1-1}) states
that $\bm{w}=\bm{v}$, we see that $\bm{b}_{1}=\bm{1}$. Now, let
$\bm{c}_{1}^{\mathsf{T}}=\left[\bm{v}^{\mathsf{T}},\bm{w}^{\mathsf{T}}\right]$
be a left eigenvector corresponding to the eigenvalue $\lambda_{1}=1$.
Then, we know that
\begin{align}
\bm{v}^{\mathsf{T}} & =\rho\bm{v}^{\mathsf{T}}\bm{A}+\bm{w}^{\mathsf{T}}\label{eq:lefteig1}\\
\bm{w}^{\mathsf{T}} & =\bm{v}^{\mathsf{T}}-\rho\bm{v}^{\mathsf{T}}\bm{D}.\label{eq:lefteig2}
\end{align}
Plugging (\ref{eq:lefteig2}) into (\ref{eq:lefteig1}) gives us that
\begin{equation}
\rho\bm{v}^{\mathsf{T}}\bm{A}+\bm{v}^{\mathsf{T}}-\rho\bm{v}^{\mathsf{T}}\bm{D}=\bm{v}^{\mathsf{T}},
\end{equation}
which implies that $\bm{v}^{\mathsf{T}}\bm{L}=\bm{0}$, and in turn
that $\bm{v}^{\mathsf{T}}=\bm{1}^{\mathsf{T}}$. Using this result
in (\ref{eq:lefteig2}) we immediately get that $\bm{c}_{1}^{\mathsf{T}}=\left[\bm{1}^{\mathsf{T}},\bm{1}^{\mathsf{T}}-\rho\bm{1}^{\mathsf{T}}\bm{D}\right]$.
\end{IEEEproof}

\subsection{\label{subsec:Proof-of-PropositionB}Proof of Proposition \ref{prop:eigv2-1}}
\begin{IEEEproof}
Denote by $\tilde{\bm{C}}^{\mathsf{T}}=\bm{B}^{-1}$ the matrix whose
rows are the scaled left eigenvectors, such that $\tilde{\bm{c}}_{i}^{\mathsf{T}}\bm{b}_{i}=1$.
In particular this means that $\tilde{\bm{c}}_{1}^{\mathsf{T}}=\bm{c}_{1}^{\mathsf{T}}/(\bm{c}_{1}^{\mathsf{T}}\bm{b}_{1})$.
Now, since 
\begin{align}
\bm{\alpha} & =\bm{B}^{-1}\bm{z}^{(0)}(\theta)\\
 & =\tilde{\bm{C}}^{\mathsf{T}}\bm{z}^{(0)}(\theta),
\end{align}
and the first row of $\tilde{\bm{C}}^{\mathsf{T}}$ is $\tilde{\bm{c}}_{1}^{\mathsf{T}}$,
then clearly $\alpha_{1}^{(0)}=\bm{c}_{1}^{\mathsf{T}}\bm{z}^{(0)}(\theta)/(\bm{c}_{1}^{\mathsf{T}}\bm{b}_{1})$.
Furthermore, since $\alpha_{1}^{(\ell)}$ is the coordinate of $\bm{z}^{(\ell)}(\theta)$
in the basis $\bm{B}$ corresponding the eigenvalue $\lambda_{1}=1$,
the part of $\bm{z}^{(\ell)}(\theta)$ (in $\bm{B}$) preserved at
each iteration is $\alpha_{1}^{(\ell)}$. Moreover, if URW-BPC converges
then from (\ref{eq:lincombeig}) and (\ref{eq:jordanlincomb}) we
observe that $\alpha_{1}$ (which is the preserved value) is the consensus
value, and since $\bm{c}_{1}^{\mathsf{T}}\bm{z}^{(0)}=2\bm{1}^{\mathsf{T}}\bm{x}^{(0)}(\theta)$
the consensus value is given by
\begin{equation}
\alpha_{1}=\frac{2}{\bm{c}_{1}^{\mathsf{T}}\bm{b}_{1}}\bm{1}^{\mathsf{T}}\bm{x}^{(0)}(\theta).
\end{equation}
\end{IEEEproof}

\subsection{\label{subsec:Proof-of-LemmaTREE}Proof of Proposition \ref{lem:tree-eigenvalues}}
\begin{IEEEproof}
The eigenvalues of $\bm{P}_{1}$ are given by the roots of the polynomial
\begin{equation}
\det\left(\bm{P}_{1}-\lambda\bm{I}_{2N}\right)\overset{\mathrm{(a)}}{=}\det\left(\lambda^{2}\bm{I}_{N}-\lambda\bm{A}+\bm{D}-\bm{I}_{N}\right)=0,\label{eq:eigofP}
\end{equation}
where the equality $\mathrm{(a})$ holds due to the four $N\times N$-blocks
of $\bm{P}_{1}-\lambda\bm{I}_{2N}$ being mutually commutative \cite[Theorem 3]{silvester2000determinants}.
For brevity, denote $\bm{\Psi}=\lambda^{2}\bm{I}_{N}-\lambda\bm{A}+\bm{D}-\bm{I}_{N}$.
Consider now the case where we add a leaf node to $\mathcal{G}$.
Without loss of generality, we assume that the leaf node is node $1$,
and its parent node is node $2$. Then, the eigenvalues of $\tilde{\bm{P}}_{1}$
are given by the roots of
\begin{align}
\det\left(\tilde{\bm{\Psi}}\right) & =\det\left(\left[\begin{array}{ccc}
\lambda^{2} & -\lambda & \bm{0}^{\mathsf{T}}\\
-\lambda & \Psi_{1,1}+1 & \bm{\Psi}_{1,2:N}\\
\bm{0} & \bm{\Psi}_{2:N,1} & \bm{\Psi}_{2:N,2:N}
\end{array}\right]\right)\\
 & =\lambda^{2}\det\left(\left[\begin{array}{cc}
\Psi_{1,1}+1 & \bm{\Psi}_{1,2:N}\\
\bm{\Psi}_{2:N,1} & \bm{\Psi}_{2:N,2:N}
\end{array}\right]\right)\\
 & +\lambda\det\left(\left[\begin{array}{cc}
-\lambda & \bm{\Psi}_{1,2:N}\\
\bm{0} & \bm{\Psi}_{2:N,2:N}
\end{array}\right]\right)\\
 & =\lambda^{2}\left(\det\left(\bm{\Psi}\right)+\det\left(\left[\begin{array}{cc}
1 & \bm{\Psi}_{1,2:N}\\
\bm{0} & \bm{\Psi}_{2:N,2:N}
\end{array}\right]\right)\right)\\
 & -\lambda^{2}\det\left(\left[\begin{array}{cc}
1 & \bm{\Psi}_{1,2:N}\\
\bm{0} & \bm{\Psi}_{2:N,2:N}
\end{array}\right]\right)\\
 & =\lambda^{2}\det\left(\bm{\Psi}\right).
\end{align}
Hence, adding a leaf node only adds two extra roots $\lambda=0$ to
the eigenvalue generating polynomial. By exchanging $\mathcal{G}$
for $\tilde{\mathcal{G}}$ and vice versa, we see that by removing
a leaf node, we remove two roots $\lambda=0$ instead. Consequently,
the nonzero eigenvalues of $\tilde{\bm{P}}_{1}$ are the same as those
of $\bm{P}_{1}$. Starting from a graph with only one node, with URW-BPC
matrix 
\begin{equation}
\bm{P}_{1}=\left[\begin{array}{cc}
0 & 1\\
1 & 0
\end{array}\right],
\end{equation}
and thus eigenvalues $\lambda_{1}=1$ and $\lambda_{2}=-1$, we see
that any tree graph with $N\ge2$ has eigenvalues $\lambda_{1}=1$,
$\lambda_{2}=-1$, and $\lambda_{i}=0$ for $i=3,\dots,2N$.
\end{IEEEproof}
\textcolor{blue}{}

\section{Proofs related to Section \ref{subsec:Regular-Graphs} }

First, we prove a few results regarding the behavior of the eigenvalues
of the URW-BPC matrix of a $k$-regular graph, with respect to the
eigenvalues of the adjacency matrix. After these useful results are
obtained, we proceed to prove the main results.
\begin{lem}
\textcolor{purple}{\label{lem:eigprops}}The roots of the polynomial
$\lambda^{2}-\mu\rho\lambda+\rho k-1=0$ are given by
\begin{equation}
\lambda_{a,b}(\mu)=\frac{1}{2}\left(\mu\rho\pm\sqrt{\mu^{2}\rho^{2}-4k\rho+4}\right).
\end{equation}
For $k\ge2$, $\mu\in\left[-k,k\right]$ and $\rho>0$ they have the
following properties:
\end{lem}
\begin{enumerate}
\item If $\lambda_{a}(\mu)$ and $\lambda_{b}(\mu)$ are complex-valued,
then $\left|\lambda_{a}(\mu)\right|=\left|\lambda_{b}(\mu)\right|=\sqrt{\left|\rho k-1\right|}$.
\item Let $\mu=k$, then $\lambda_{b}\left(k\right)=1$ and $\lambda_{a}\left(k\right)=\rho k-1$. 
\item $\left|\lambda_{a}(\mu)\right|=\left|\lambda_{b}(-\mu)\right|$.
\item $\left|\lambda_{a}\left(\mu\right)\right|$ is a nondecreasing function
of $\mu$; $\left|\lambda_{b}(\mu)\right|$ is a nonincreasing function
of $\mu$.
\end{enumerate}
\begin{IEEEproof}
When the roots $\lambda_{a}(\mu),\lambda_{b}(\mu)$ are complex-valued,
the squared absolute values are given by
\begin{align}
\left|\lambda_{a,b}(\mu)\right|^{2} & =\frac{1}{4}\left|\mu\rho\pm i\sqrt{\left|4k\rho-4\right|-\mu^{2}\rho^{2}}\right|^{2}\\
 & =\frac{1}{4}\left(\mu^{2}\rho^{2}+\left|4k\rho-4\right|-\mu^{2}\rho^{2}\right)\\
 & =\left|k\rho-1\right|,
\end{align}
where $i=\sqrt{-1}$. Hence, for complex-valued $\lambda_{a,b}(\mu)$
we have $\left|\lambda_{a}(\mu)\right|=\left|\lambda_{b}(\mu)\right|=\sqrt{\left|k\rho-1\right|}$.

For property (ii), suppose first that $k\rho\ge2$. We plug in $\mu=k$
\begin{align}
\left|\lambda_{a}\left(k\right)\right| & =\frac{1}{2}\left|k\rho+\sqrt{k^{2}\rho^{2}-4k\rho+4}\right|\\
 & =\frac{1}{2}\left|k\rho+\left(k\rho-2\right)\right|\\
 & =k\rho-1.
\end{align}
\begin{align}
\left|\lambda_{b}\left(k\right)\right| & =\frac{1}{2}\left|k\rho-k\rho+2\right|\\
 & =1.
\end{align}
For $k\rho<2$, the roots are interchanged.

For (iii) we have that
\begin{align}
\left|\lambda_{b}\left(-\mu\right)\right| & =\left|-\mu\rho-\sqrt{\mu^{2}\rho^{2}-4k\rho+4}\right|\\
 & =\left|-\left(\mu\rho+\sqrt{\mu^{2}\rho^{2}-4k\rho+4}\right)\right|\\
 & =\left|\mu\rho+\sqrt{\mu^{2}\rho^{2}-4k\rho+4}\right|\\
 & =\left|\lambda_{a}\left(\mu\right)\right|.
\end{align}

To show (iv), we focus on the case when $\lambda_{a}(\mu)$ is real,
since for complex $\lambda_{a}(\mu)$, $\left|\lambda_{a}(\mu)\right|$
is constant in $\mu$. We check that the derivative of $\lambda_{a}^{2}\left(\mu\right)$
wrt. $\mu$ is positive, and since $\lambda_{a}\left(\mu\right)$
is assumed to be real this holds for $\left|\lambda_{a}\left(\mu\right)\right|$
as well. The derivative of $\lambda_{a}^{2}\left(\mu\right)$ wrt.
$\mu$ is given by
\begin{align}
\frac{\partial}{\partial\mu}\lambda_{a}^{2}\left(\mu\right) & =2\lambda_{a}\left(\mu\right)\frac{\partial}{\partial\mu}\lambda_{a}\left(\mu\right)\\
 & =\frac{2\lambda_{a}^{2}\left(\mu\right)\rho}{\sqrt{\mu^{2}\rho^{2}-4k\rho+4}}.
\end{align}
Since $\rho>0$, the derivative of $\lambda_{a}^{2}\left(\mu\right)$
is clearly positive, and thus so is the derivative of $\left|\lambda_{a}\left(\mu\right)\right|$.
We conclude that $\left|\lambda_{a}(\mu)\right|$ is nondecreasing
in $\mu$, and due to (iii) that $\left|\lambda_{b}(\mu)\right|$
is nonincreasing in $\mu$. Note that the functions are not necessarily
monotonic since they are constant for complex eigenvalues.\textcolor{blue}{}
\end{IEEEproof}

\subsection{Proof of Lemma \ref{lem:secondlargest} \label{subsec:secondlargets-proof}}
\begin{IEEEproof}
Let $\lambda\neq0$ be an eigenvalue of $\bm{P}_{\rho}$ and $\left[\bm{v}^{\mathsf{T}},\bm{w}^{\mathsf{T}}\right]^{\mathsf{T}}\neq\bm{0}$
the corresponding eigenvector. By definition, it holds that

\begin{align}
\lambda\bm{v} & =\rho\bm{A}\bm{v}+(\bm{I}_{N}-\rho\bm{D})\bm{w}\label{eq:kregeig1-1}\\
\lambda\bm{w} & =\bm{v}.
\end{align}
Substituting $\bm{v}$ by $\lambda\bm{w}$ in (\ref{eq:kregeig1-1})
gives

\begin{equation}
\lambda^{2}\bm{w}=\rho\bm{A}\lambda\bm{w}+(\bm{I}_{N}-\rho\bm{D})\bm{w}.
\end{equation}
Since $\mathcal{G}$ is a $k$-regular graph, we have that $\bm{D}=k\bm{I}_{N}$,
and since $\lambda\neq0$, this is equivalent to
\begin{equation}
\bm{Aw}=\frac{\lambda^{2}+\rho k-1}{\rho\lambda}\bm{w}.\label{eq:kregeig2}
\end{equation}
This means that the eigenvalues of $\bm{P}_{\rho}$ are given by the
roots of the polynomial
\begin{equation}
\lambda=\frac{1}{2}\left(\mu\rho\pm\sqrt{\mu^{2}\rho^{2}-4k\rho+4}\right),
\end{equation}
where $\mu$ is an eigenvalue of $\bm{A}$ with eigenvector $\bm{w}$.
Suppose that $\mu_{i}>0$. Then, Lemma \ref{lem:eigprops}, which
says that $\lambda_{a}\left(\mu\right)$ is a nondecreasing function
in $\mu$, implies that $\lambda_{i}=\lambda_{a}\left(\mu_{i}\right)$
and thus its magnitude is given by
\begin{equation}
\left|\lambda_{i}\right|=\frac{1}{2}\left|\mu_{i}\rho+\sqrt{\mu_{i}\rho-4k\rho+4}\right|.
\end{equation}
Since $\left|\lambda_{a}\left(\mu\right)\right|=\left|\lambda_{b}\left(-\mu\right)\right|$
by Lemma \ref{lem:eigprops}, we observe that we achieve the same
result should $\mu_{i}<0$.
\end{IEEEproof}

\subsection{Proof of Theorem \ref{lem:kconvergence}\label{subsec:kconvergence-proof}}
\begin{IEEEproof}
Let $\tilde{\mu}=\max_{i}\left|\mu_{i}\right|$ for $i$ such that
$\left|\mu_{i}\right|<k$. Note that for a non-bipartite $\mathcal{G}$,
$\tilde{\mu}=\mu_{2}$, whereas for a bipartite $\mathcal{G}$ we
have that $\mu_{2}=-k$, and hence $\tilde{\mu}=\mu_{3}$. Since $\left|\lambda_{a}\left(\mu\right)\right|$
is a nondecreasing function in $\mu$ (due to Lemma \ref{lem:eigprops})
and $\tilde{\mu}<k$, and if $k\rho<2$ we have that
\begin{align}
|\tilde{\lambda}| & =\frac{1}{2}\left|\tilde{\mu}\rho+\sqrt{\tilde{\mu}^{2}\rho^{2}-4k\rho+4}\right|\\
 & <\frac{1}{2}\left|k\rho+\sqrt{k^{2}\rho^{2}-4k\rho+4}\right|\\
 & =\frac{1}{2}\left|k\rho+\left(2-k\rho\right)\right|\\
 & =1.
\end{align}
Hence, $\left|\lambda_{i}\right|<1$ for $i=2,\dots,2N$. On the other
hand, if $k\rho>2$, we have that
\begin{align}
\left|\lambda_{1}\right| & =\frac{1}{2}\left|k\rho+\sqrt{k^{2}\rho^{2}-4k\rho+4}\right|\\
 & =\frac{1}{2}\left|k\rho+k\rho-2\right|\\
 & =k\rho-1\\
 & >1.
\end{align}
So, in that case URW-BPC is not convergent. In particular for $\rho=2/k$
we have that
\begin{align}
\lambda & =\frac{1}{2}\left(\mu\frac{2}{k}\pm\sqrt{\mu^{2}\frac{4}{k^{2}}-4}\right),
\end{align}
so all eigenvalues are complex-valued except the ones generated from
$\mu=k$ or $\mu=-k$ (the smallest eigenvalue of $\bm{A}$ for bipartite
$\mathcal{G}$ is $\mu=-k$ \cite{lovasz2007eigenvalues}), which
are equal to $\lambda=1$ or $\lambda=-1$. Thus, using property (i)
we find that $\left|\lambda_{i}\right|=1$ for all $i=1,\dots,2N$.
Moreover, for $\rho=0$ we clearly see that $\left|\lambda_{i}\right|=1$
for all $i=1,\dots,2N$. For $\rho<0$ it is obvious that $\left|\lambda_{1}\right|>1$.

Using the results from Propositions \ref{prop:eigv1} and \ref{prop:eigv2-1},
and that the sum of the degrees for undirected $k$-regular graphs
is $\sum_{i=1}^{N}D_{ii}=Nk$, the consensus value $\alpha_{1}$ is
given
\begin{equation}
\alpha_{1}=\frac{2}{2N-\rho Nk}\sum_{m=1}^{N}x_{m}^{(0)}(\theta).
\end{equation}
\end{IEEEproof}

\subsection{Proof of Theorem \ref{thm:optrho}\label{subsec:Proof-of-TheoremCR}}

\textcolor{blue}{}

\begin{IEEEproof}
We want to find the $\rho$ that minimizes the magnitude of the largest
eigenvalue inside the unit circle, i.e., $|\tilde{\lambda}|$. Let
$\tilde{\mu}$ be the eigenvalue of $\bm{A}$ that generates $\tilde{\lambda}$.
Then, we minimize $|\tilde{\lambda}|$ by
\begin{equation}
\min_{\rho\in(0,1]}\frac{1}{2}\left|\tilde{\mu}\rho+\sqrt{\tilde{\mu}^{2}\rho^{2}-4k\rho+4}\right|,\label{eq:optrhoabs}
\end{equation}
First we get the roots with respect to $\rho$ of the polynomial under
the square-root
\begin{align}
\rho & =\frac{2k}{\tilde{\mu}^{2}}\pm\sqrt{\frac{4k^{2}}{\tilde{\mu}^{4}}-4}\\
 & =\frac{2}{\tilde{\mu}^{2}}\left(k\pm\sqrt{k^{2}-\tilde{\mu}^{2}}\right).
\end{align}
Since $\sqrt{k^{2}-\tilde{\mu}^{2}}>0$ and $k>\sqrt{k^{2}-\tilde{\mu}^{2}}$,
we see that the smallest $\rho$ is given by
\begin{equation}
\rho^{\star}=\frac{2}{\tilde{\mu}^{2}}\left(k-\sqrt{k^{2}-\tilde{\mu}^{2}}\right).
\end{equation}
This value of $\rho$ will make the second term inside the absolute
value in (\ref{eq:optrhoabs}) equal to zero, yielding
\begin{equation}
|\tilde{\lambda}|=\left|\tilde{\mu}\rho^{\star}\right|.\label{eq:abslam}
\end{equation}
However, it is still not clear that this is the global minimum, since
there is a linear term in the expression too. \textcolor{black}{First,
since $\tilde{\mu}$ is positive, $\rho>\rho^{\star}$ cannot give
smaller $|\tilde{\lambda}|$ than the one given by $\rho^{\star}$.}
But, there might be a $\rho<\rho^{\star}$ that gives a smaller $|\tilde{\lambda}|$.
So, consider using $\rho_{\epsilon}=\rho^{\star}-\epsilon$, where
$\epsilon>0$. Then we get
\begin{equation}
|\tilde{\lambda}_{\epsilon}|=\frac{1}{2}\left|\tilde{\mu}\rho^{\star}-\tilde{\mu}\epsilon+\sqrt{\tilde{\mu}^{2}\epsilon^{2}+4\epsilon\sqrt{k^{2}-\tilde{\mu}^{2}}}\right|.
\end{equation}
Since $4\epsilon\sqrt{k^{2}-\tilde{\mu}^{2}}>0$, we have that 
\begin{equation}
\sqrt{\tilde{\mu}^{2}\epsilon^{2}+4\epsilon\sqrt{k^{2}-\tilde{\mu}^{2}}}>\tilde{\mu}\epsilon,
\end{equation}
and hence $|\tilde{\lambda}_{\epsilon}|>|\tilde{\lambda}|$. Consequently,
the optimal $\rho$ is
\begin{equation}
\rho_{\mathrm{opt}}=\frac{2}{\tilde{\mu}^{2}}\left(k-\sqrt{k^{2}-\tilde{\mu}^{2}}\right),
\end{equation}
and, plugging this value into (\ref{eq:abslam}) gives
\begin{align}
|\tilde{\lambda}| & =\frac{1}{2}\left|\frac{2}{\tilde{\mu}}\left(k-\sqrt{k^{2}-\tilde{\mu}^{2}}\right)\right|\\
 & =\left|\frac{1}{\tilde{\mu}}\left(k-\sqrt{k^{2}-\tilde{\mu}^{2}}\right)\right|.
\end{align}
\end{IEEEproof}

\subsection{\textcolor{black}{Proof of Proposition \ref{prop:bipartiteeig} \label{subsec:bipartiteeig-proof}}}
\begin{IEEEproof}
\textcolor{black}{Denote by $\alpha(\lambda)$ and $\gamma(\lambda)$
the algebraic and geometric multiplicities of an eigenvalue $\lambda$
of a URW-BPC matrix $\bm{P}_{\rho}$. Suppose that $\alpha(\lambda)\neq\gamma(\lambda)$,
i.e., $\alpha(\lambda)>\gamma(\lambda)$. Then, there exist vectors
$\bm{v}$ and $\bm{w}$ such that
\begin{equation}
\bm{P}_{\rho}\left[\begin{array}{c}
\bm{v}\\
\bm{w}
\end{array}\right]=\lambda\left[\begin{array}{c}
\bm{v}\\
\bm{w}
\end{array}\right]+\left[\begin{array}{c}
\tilde{\bm{v}}\\
\tilde{\bm{w}}
\end{array}\right],\label{eq:generalizedeig}
\end{equation}
where $[\tilde{\bm{v}}^{\mathsf{T}},\tilde{\bm{w}}^{\mathsf{T}}]^{\mathsf{T}}$
is an eigenvector of $\bm{P}_{\rho}$ with eigenvalue $\lambda$.
As established in (\ref{eq:kregeig1-1})\textendash (\ref{eq:kregeig2}),
if $\lambda\neq0$
\begin{equation}
\left[\begin{array}{c}
\tilde{\bm{v}}\\
\tilde{\bm{w}}
\end{array}\right]=\left[\begin{array}{c}
\lambda\bm{z}\\
\bm{z}
\end{array}\right],\label{eq:Peigvec}
\end{equation}
for some $\bm{z}$ such that
\begin{align}
\bm{A}\bm{z} & =\mu\bm{z}\label{eq:Aeigvec}\\
\mu & =\frac{\lambda^{2}+\rho k-1}{\rho\lambda}.\label{eq:mudef}
\end{align}
Using (\ref{eq:Peigvec}) in (\ref{eq:generalizedeig}), we get that
\begin{align}
\rho\bm{A}\bm{v}+(1-\rho k)\bm{w} & =\lambda\bm{v}+\lambda\bm{z}\label{eq:geneigexp1}\\
\bm{v} & =\lambda\bm{w}+\bm{z}.
\end{align}
Substituting $\bm{v}$ in (\ref{eq:geneigexp1}), we have
\begin{equation}
\rho\bm{A}\lambda\bm{w}+\rho\bm{A}\bm{z}+(1-\rho k)\bm{w}=\lambda^{2}\bm{w}+2\lambda\bm{z},
\end{equation}
which in turn, using (\ref{eq:Aeigvec}), becomes
\begin{equation}
\rho\lambda\bm{A}\bm{w}=-\rho\mu\bm{z}+(\rho k-1)\bm{w}+\lambda^{2}\bm{w}+2\lambda\bm{z}.
\end{equation}
Rearranging the terms, we have that
\begin{align}
\bm{A}\bm{w} & =\frac{\lambda^{2}+\rho k-1}{\rho\lambda}\bm{w}+\frac{2\lambda-\rho\mu}{\rho\lambda}\bm{z}\\
 & =\mu\bm{w}+\frac{2\lambda-\rho\mu}{\rho\lambda}\bm{z}.
\end{align}
Left-multiplying by $\bm{z}^{\mathsf{T}}$ and using the symmetry
of $\bm{A}$ (so that $\bm{z}^{\mathsf{T}}\bm{A}=\bm{z}^{\mathsf{T}}\mu$),
we get
\begin{equation}
\mu\bm{z}^{\mathsf{T}}\bm{w}=\mu\bm{z}^{\mathsf{T}}\bm{w}+\frac{2\lambda-\rho\mu}{\rho\text{\ensuremath{\lambda}}}\left\Vert \bm{z}\right\Vert ^{2}.
\end{equation}
This implies that $(2\lambda-\rho\mu)/\rho\lambda=0$, and thus that
\begin{align}
2\lambda & =\rho\mu\\
 & =\frac{\lambda^{2}+\rho k-1}{\lambda}.
\end{align}
Hence, we have that $\lambda^{2}=\rho k-1$. For $\lambda=\pm1$,
this implies that $\rho k=2$. But, $\rho\in(0,2/k)$, hence the original
claim is false. We conclude that $\alpha(\lambda)=\gamma(\lambda)$
for $\lambda=\pm1$.}
\end{IEEEproof}

\textcolor{magenta}{}

\bibliographystyle{IEEEtran}
\bibliography{BibBPC}

\begin{thebibliography}{10}
\providecommand{\url}[1]{#1}
\csname url@samestyle\endcsname
\providecommand{\newblock}{\relax}
\providecommand{\bibinfo}[2]{#2}
\providecommand{\BIBentrySTDinterwordspacing}{\spaceskip=0pt\relax}
\providecommand{\BIBentryALTinterwordstretchfactor}{4}
\providecommand{\BIBentryALTinterwordspacing}{\spaceskip=\fontdimen2\font plus
\BIBentryALTinterwordstretchfactor\fontdimen3\font minus
  \fontdimen4\font\relax}
\providecommand{\BIBforeignlanguage}[2]{{%
\expandafter\ifx\csname l@#1\endcsname\relax
\typeout{** WARNING: IEEEtran.bst: No hyphenation pattern has been}%
\typeout{** loaded for the language `#1'. Using the pattern for}%
\typeout{** the default language instead.}%
\else
\language=\csname l@#1\endcsname
\fi
#2}}
\providecommand{\BIBdecl}{\relax}
\BIBdecl

\bibitem{pearl1986fusion}
J.~Pearl, ``Fusion, propagation, and structuring in belief networks,''
  \emph{Artificial intelligence}, vol.~29, no.~3, pp. 241--288, Sep. 1986.

\bibitem{kschischang2001factor}
F.~R. Kschischang, B.~J. Frey, and H.-A. Loeliger, ``Factor graphs and the
  sum-product algorithm,'' \emph{IEEE Transactions on Information Theory},
  vol.~47, no.~2, pp. 498--519, Feb. 2001.

\bibitem{yedidia2005constructing}
J.~S. Yedidia, W.~T. Freeman, and Y.~Weiss, ``Constructing free-energy
  approximations and generalized belief propagation algorithms,'' \emph{IEEE
  Transactions on Information Theory}, vol.~51, no.~7, pp. 2282--2312, Jul.
  2005.

\bibitem{sun2003stereo}
J.~Sun, N.-N. Zheng, and H.-Y. Shum, ``Stereo matching using belief
  propagation,'' \emph{IEEE Transactions on Pattern Analysis and Machine
  Intelligence}, vol.~25, no.~7, pp. 787--800, Jul. 2003.

\bibitem{moallemi2009convergence}
C.~C. Moallemi and B.~Van~Roy, ``Convergence of min-sum message passing for
  quadratic optimization,'' \emph{IEEE Transactions on Information Theory},
  vol.~55, no.~5, pp. 2413--2423, May 2009.

\bibitem{chamley2013models}
C.~Chamley, A.~Scaglione, and L.~Li, ``Models for the diffusion of beliefs in
  social networks: An overview,'' \emph{IEEE Signal Processing Magazine},
  vol.~30, no.~3, pp. 16--29, May 2013.

\bibitem{kabashima2003cdma}
Y.~Kabashima, ``A {CDMA} multiuser detection algorithm on the basis of belief
  propagation,'' \emph{Journal of Physics A: Mathematical and General},
  vol.~36, no.~43, p. 11111, Oct. 2003.

\bibitem{wymeersch2009cooperative}
H.~Wymeersch, J.~Lien, and M.~Z. Win, ``Cooperative localization in wireless
  networks,'' \emph{Proceedings of the IEEE}, vol.~97, no.~2, pp. 427--450,
  Feb. 2009.

\bibitem{fossorier1999reduced}
M.~P. Fossorier, M.~Mihaljevi\'{c}, and H.~Imai, ``Reduced complexity iterative
  decoding of low-density parity check codes based on belief propagation,''
  \emph{IEEE Transactions on Communications}, vol.~47, no.~5, pp. 673--680, May
  1999.

\bibitem{mceliece1998turbo}
R.~J. McEliece, D.~J.~C. MacKay, and J.-F. Cheng, ``Turbo decoding as an
  instance of {Pearl's} "belief propagation" algorithm,'' \emph{IEEE Journal on
  Selected Areas in Communications}, vol.~16, no.~2, pp. 140--152, Feb. 1998.

\bibitem{zarrin2008belief}
S.~Zarrin and T.~J. Lim, ``Belief propagation on factor graphs for cooperative
  spectrum sensing in cognitive radio,'' in \emph{Proceedings of the 3rd IEEE
  Symposium on New Frontiers in Dynamic Spectrum Access Networks}, Oct. 2008.

\bibitem{meyer2012simultaneous}
F.~Meyer, E.~Riegler, O.~Hlinka, and F.~Hlawatsch, ``Simultaneous distributed
  sensor self-localization and target tracking using belief propagation and
  likelihood consensus,'' in \emph{Conference Record of the 46th Asilomar
  Conference on Signals, Systems and Computers}, Mar. 2013, pp. 1212--1216.

\bibitem{etzlinger2014cooperative}
B.~Etzlinger, H.~Wymeersch, and A.~Springer, ``Cooperative synchronization in
  wireless networks.'' \emph{IEEE Transactions on Signal Processing}, vol.~62,
  no.~11, pp. 2837--2849, Jun. 2014.

\bibitem{zhong2005ldgm}
W.~Zhong and J.~Garcia-Frias, ``{LDGM} codes for channel coding and joint
  source-channel coding of correlated sources,'' \emph{EURASIP Journal on
  Applied Signal Processing}, vol. 2005, pp. 942--953, Jan. 2005.

\bibitem{zhang2011belief}
Z.~Zhang, Z.~Han, H.~Li, D.~Yang, and C.~Pei, ``Belief propagation based
  cooperative compressed spectrum sensing in wideband cognitive radio
  networks,'' \emph{IEEE Transactions on Wireless Communications}, vol.~10,
  no.~9, pp. 3020--3031, Jul. 2011.

\bibitem{wainwright2003tree}
M.~J. Wainwright, T.~S. Jaakkola, and A.~S. Willsky, ``Tree-reweighted belief
  propagation algorithms and approximate ml estimation by pseudo-moment
  matching,'' in \emph{Proceedings of the 9th International Workshop on
  Artificial Intelligence and Statistics}, Jan. 2003.

\bibitem{wainwright2005new}
------, ``A new class of upper bounds on the log partition function,''
  \emph{IEEE Transactions on Information Theory}, vol.~51, no.~7, pp.
  2313--2335, Jun. 2005.

\bibitem{wymeersch2011uniformly}
H.~Wymeersch, F.~Penna, and V.~Savi{\'c}, ``Uniformly reweighted belief
  propagation: A factor graph approach,'' in \emph{Proceedings of the IEEE
  International Symposium on Information Theory}, Oct. 2011, pp. 2000--2004.

\bibitem{wymeersch2012uniformly}
------, ``Uniformly reweighted belief propagation for estimation and detection
  in wireless networks,'' \emph{IEEE Transactions on Wireless Communications},
  vol.~11, no.~4, pp. 1587--1595, Feb. 2012.

\bibitem{liu2012knowledge}
J.~Liu and R.~C. de~Lamare, ``Knowledge-aided reweighted belief propagation
  decoding for regular and irregular ldpc codes with short blocks,'' in
  \emph{Proceedings of the International Symposium on Wireless Communication
  Systems}, Oct. 2012, pp. 984--988.

\bibitem{liu2012low}
------, ``Low-latency reweighted belief propagation decoding for ldpc codes,''
  \emph{IEEE Communications Letters}, vol.~16, no.~10, pp. 1660--1663, Aug.
  2012.

\bibitem{olfati2007con}
R.~Olfati-Saber, J.~A. Fax, and R.~M. Murray, ``Consensus and cooperation in
  networked multi-agent systems,'' \emph{Proceedings of the IEEE}, vol.~95,
  no.~1, pp. 215--233, Mar. 2007.

\bibitem{xiao2006distributed}
\BIBentryALTinterwordspacing
L.~Xiao, S.~Boyd, and S.~Lall, ``Distributed average consensus with
  time-varying metropolis weights,'' 2006. [Online]. Available:
  \url{http://web.stanford.edu/~boyd/papers/pdf/avg_metropolis.pdf}
\BIBentrySTDinterwordspacing

\bibitem{dai2007consensus}
H.~Dai and Y.~Zhang, ``Consensus estimation via belief propagation,'' in
  \emph{Proceedings of the 41st Annual Conference on Information Sciences and
  Systems}, Sep. 2007, pp. 277--281.

\bibitem{lovasz2007eigenvalues}
\BIBentryALTinterwordspacing
L.~Lov\'{a}sz, ``Eigenvalues of graphs,'' 2007. [Online]. Available:
  \url{http://www.cs.elte.hu/~lovasz/eigenvals-x.pdf}
\BIBentrySTDinterwordspacing

\bibitem{mckay1981expected}
B.~D. McKay, ``The expected eigenvalue distribution of a large regular graph,''
  \emph{Linear Algebra and its Applications}, vol.~40, pp. 203--216, Oct. 1981.

\bibitem{savic2010indoor}
V.~Savi{\'c}, A.~Poblaci{\'o}n, S.~Zazo, and M.~Garc{\'\i}a, ``Indoor
  positioning using nonparametric belief propagation based on spanning trees,''
  \emph{EURASIP Journal on Wireless Communications and Networking}, vol. 2010,
  no.~1, Jul. 2010.

\bibitem{farkas2001spectra}
I.~J. Farkas, I.~Der{\'e}nyi, A.-L. Barab{\'a}si, and T.~Vicsek, ``Spectra of
  real-world graphs: Beyond the semicircle law,'' \emph{Physical Review E},
  vol.~64, no.~2, Jul. 2001.

\bibitem{silvester2000determinants}
J.~R. Silvester, ``Determinants of block matrices,'' \emph{The Mathematical
  Gazette}, vol.~84, no. 501, pp. 460--467, Nov. 2000.

\end{thebibliography}

\end{document}